# Superelastic softening of ferroelastic multidomain crystals

W. SCHRANZ[1,*], H. KABELKA[1] and A. TRÖSTER[2]

[1] University of Vienna, Faculty of Physics, Physics of Functional Materials, Boltzmanngasse 5, A-1090 Wien, Austria

[2] Johannes Gutenberg Universität Mainz, Staudingerweg 7, D-55099 Mainz, Germany

Many proper and improper ferroelastic materials display (at sufficiently low measurement frequencies) a huge elastic softening below $T_c$. This giant elastic softening, which can be suppressed with uniaxial stress, is caused by domain wall motion. Here we shortly review our results on frequency and temperature dependent elastic measurements of some perovskites which exhibit improper ferroelastic phase transitions. We also present a new model - based on Landau-Ginzburg theory including long range interaction of needle shaped ferroelastic domains - which describes superelastic softening observed in some of the perovskite systems very well. We also show, how the theory can be extended to describe proper ferroelastic materials and apply the theory to describe the elastic behaviour of the proper ferroelastic material $La_{1-x}Nd_xP_5O_{14}$ (LNPP).

## 1. Introduction

Ferroelastic materials are an important subset of ferroic materials[1,2]. Depending on the type of coupling between the primary order parameter $\eta$ and the strain variables $\varepsilon_i$ (i=1,..,6 in Voight notation) the ferroelastic state can be classified as proper, pseudo proper, improper or co-elastic[3].

Very often ferroelastic materials consist of a large number of elastic domains which are separated by domain boundaries. Due to mechanical compatibility between adjacent domains ferroelastic domain walls are generally well oriented and planar. However, very often needle or dagger shaped domains appear[3] as will be discussed in some detail below. In the presence of an applied external stress these domain boundaries can move, which can drastically influence the macroscopic elastic properties of the material.

---

[*] Corresponding author, E-mail: wilfried.schranz@univie.ac.at

Understanding the macroscopic behaviour of multidomain crystals is important for technological applications as well as for a better understanding of the seismic properties of our earth. E.g. domain wall motion may well influence the low frequency elastic and anelastic behaviour of mantle minerals at seismic frequencies (1-20 Hz).

In recent years we have performed low frequency (0.1 Hz - 100 Hz) elastic measurements in quite a number of perovskite structured materials ($SrTiO_3$, $KMnF_3$, etc.), measuring usually a huge softening in the low symmetry improper ferroelastic phases[4,5,6]. Very similar results have been obtained for $LaAlO_3$[7] and $Sr_xCa_{1-x}TiO_3$ [8] and with mechanical shear measurements on $SrTiO_3$ [9,10] and $Hg_2Cl_2$[11]. Calculation of the response of a ferroelastic multidomain crystal is hampered by the fact, that planar ferroelastic domain walls are usually metastable objects, i.e. no equilibrium free energy can be constructed for an array of parallel planar ferroelastic domain walls. This is in contrast to ferroelectric or ferromagnetic domains, where the competition between the domain wall energy and the long range depolarization[12] or demagnetization[13] field leads to a stable domain pattern. There the equilibrium free energy can be used to calculate the average number of domains (i.e. domain width), its change with temperature and/or applied external field as well as the corresponding macroscopic susceptibilities. All these nice features seemed to be absent for ferroelastic domains and researchers have mainly focussed on first order phase front-[14,15] and substrate-[16] stabilized ferroelastic domain arrays. However, in a previous work[17] Torrés, *et al.* have shown, that at the end of needle shaped ferroelastic domain walls long range elastic stress fields are produced, which are reminiscent of the stray fields in ferroelectric or ferromagnetic materials. Including the corresponding interaction term in the Landau-Ginzburg free energy one can calculate the elastic response of the ferroelastic multidomain crystal. Here we summarize the main results for *improper ferroelastic perovskites*[18] and compare them with the multidomain elastic behaviour of *proper ferroelastics*, which turns out to be very different.

**2. Experimental**

For the low frequency elastic measurements a Dynamical Mechanical Analyzer (DMA7-Perkin Elmer) is used. The samples are exposed to a given static force $F_{stat}$ (tunable between 1 mN and 2500 mN with a precision of 1 mN) modulated by a dynamic force $F_{dyn}$ of chosen amplitude and frequency (0.1-50Hz). The amplitude $u$ and the phase shift $\delta$ of the resulting elastic response of a sample are registered *via* inductive coupling with a resolution of $\Delta u \approx 10$nm and $\Delta\delta \approx 0.06°$. The knowledge of $u$ and $\delta$ allows the determination of both real and imaginary parts of the inverse complex elastic compliance (Young's modulus Y*). The measurements have been performed by the parallel-plate stress (PPS) or the three-point bending (TPB) method (see Fig.1). In PPS geometry the complex Young's modulus in the direction $\vec{q}$ of the applied force $Y^*(\vec{q}) = Y'(\vec{q}) + iY''(\vec{q})$ is determined as:

$$Y^*(\vec{q}) = \frac{F_{dyn}}{u} \frac{h}{A} \exp(i\delta) \qquad (1)$$

where h and A represent the sample thickness and area, respectively. In our studies we used samples with typical dimensions: $A \approx 1\text{-}4$ mm$^2$ and $h \approx 3\text{-}5$ mm.

In TPB geometry – where $\vec{p}$ is perpendicular to the applied force and pointing along the long axis of the sample bar – one gets

$$Y^*(\vec{p}) \approx \frac{F_{dyn}}{u} \frac{L^3}{4bh^3} \exp(i\delta) \qquad (2)$$

In this geometry one usually uses thin bars of $h \approx 0.2\text{-}0.5$ mm, $L > 5$ mm and $b \approx 2$ mm. The absolute accuracy of such measurements is usually not better than 20%, whereas the relative accuracy of the DMA method is within 0.2-1%. For more details of the method and its application to phase transitions, glass freezing, etc. see e.g.[19,20].

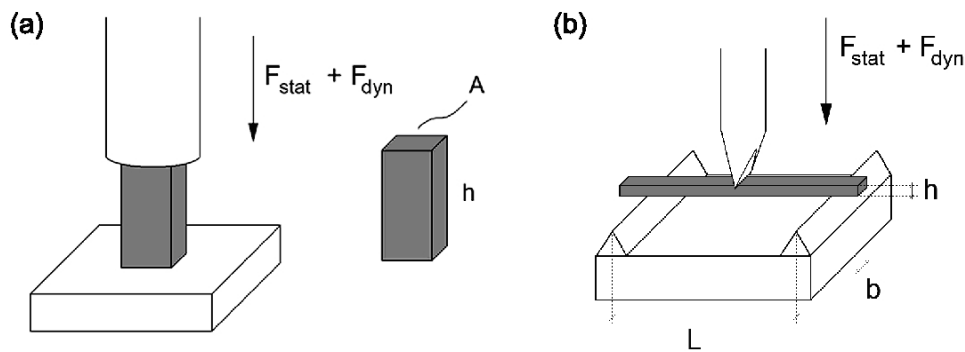

**Figure 1**. Dynamic mechanical analysis (DMA) measurements in (a) parallel plate (PPS) and (b) three point bending (TPB) geometry.

## 3. Modelling the domain wall contribution to the elastic susceptibility

### 3.1. Improper ferroelastic phase transitions

Many perovskites, e.g. $SrTiO_3$, $KMnF_3$, $LaAlO_3$ exhibit improper ferroelastic phase transitions, where the order parameter η couples quadratically with the corresponding strain ε, i.e. a term ~$η^2ε$ exists in the Landau free energy expansion. This type of coupling leads to the so called Landau-Khalatnikov (LK) contribution to the elastic anomaly, which is manifested in a negative jump, or dip like anomaly at $T_c$ in the real part and a peak in the imaginary part of the complex elastic constant [21]. Meanwhile these LK-contributions have been measured in many different materials including $SrTiO_3$[22], $KMnF_3$[6], $C_{60}$[23], $KSCN$[24], etc. However, after extensive measurements of mainly perovskite structured materials[4-8] it became obvious to us, that their elastic behaviour measured in the Hz region differs drastically from their ultrasonic behaviour at MHz frequencies. The high frequency elastic behaviour of these perovskite crystals is well described by Landau theory (i.e. LK-type anomaly), whereas their low frequency response below $T_c$ is dominated by domain wall motion. It turned out, that in spite of the complicated domain structure appearing in the improper ferroelastic phase of these crystals the domain wall response to the elastic compliance $\Delta S^{DW}$ showed a rather universal behaviour: a huge softening of $Y'$ (equivalent to increase of the compliance $S'$) below $T_c$, whose temperature dependence is proportional to the square of the order parameter, i.e.

$$\Delta S^{DW} \propto \eta^2(T) \qquad (3)$$

It should be mentioned that since the temperature dependencies of the order parameters for $SrTiO_3$, $KMnF_3$[25] and $LaAlO_3$[26] have been measured very accurately, there is no freedom in the fitting procedure using Eq.(3). In the following we will show, how Eq.(3) – which was at first empirically found for $SrTiO_3$[5] – can be justified by a theoretical model. Inspecting the real domain structure of the considered perovskites one finds, that they all show a similar domain pattern in the improper ferroelastic phase, i.e. rather long needle shaped domains, oriented in prominent directions due to mechanical compatibility between adjacent ferroelastic domains (Fig.2).

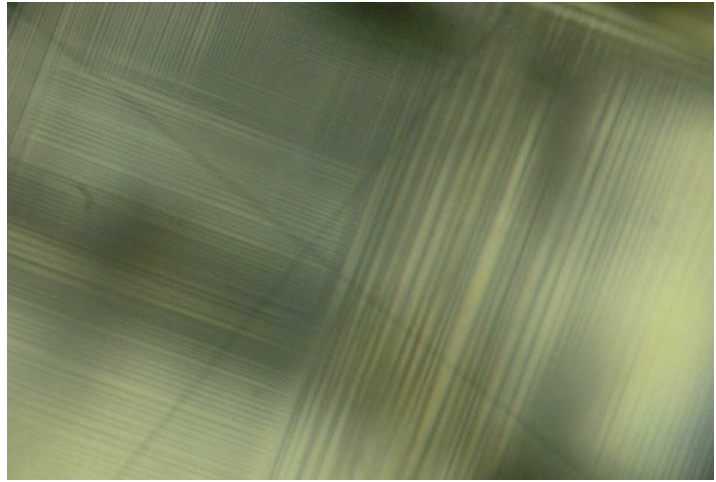

**Figure 2**. Pattern of needle shaped ferroelastic domains in $KMn_{0.983}Ca_{0.017}F_3$ observed at T=183 K with a polarizing microscope.

The origin and physical properties of such needle shaped domains were subjects of many studies[27,28,29,30,31,32,33]. One possibility to understand the origin of needle shaped domains is as follows: Due to the appearance of a spontaneous strain $\varepsilon_s$ in the ferroelastic phase the angle $\phi$ between equivalent domain walls can change to $\phi \pm \varepsilon_s$ (Fig.3). The corresponding lattice mismatch can be described by wedge disclinations located at the domain wall intersections. Around the left corner of Fig.3 there is excess of matter, whereas on the right side lack of matter appears, which implies, that the two corners (indicated as + and -) attract each other, until they form a needle tip.

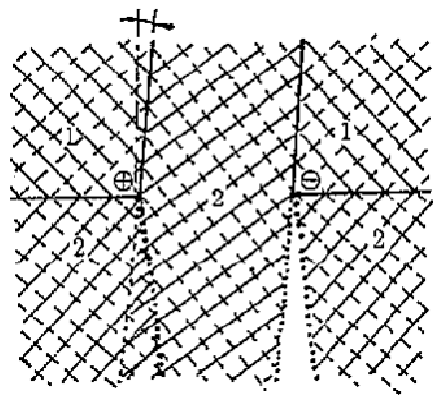

**Figure 3**. Schematic representation of the lattice mismatch of two almost perpendicular ($90° \pm \varepsilon_s$) domain walls (from Ref.28).

These needle tips deviate from the coherent domain wall orientation, which then can be described by the appearance of dislocations. For rare-earth monoclinic sesquioxides ($Ln_2O_3$-B) a high resolution electron microscopy study has indeed shown[34] that the regions near the needle tips which deviate from the coherent orientation are made of a succession of short coherent boundaries separated by small steps acting as dislocations.

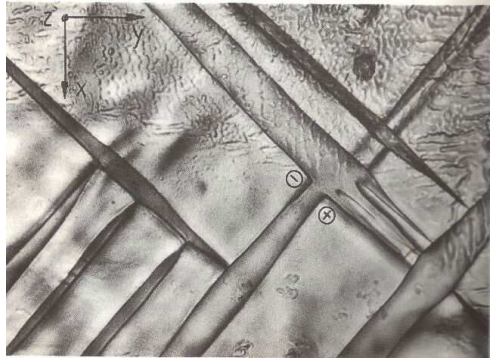

**Figure 4**. Optical micrograph of domain wall pattern of the improper ferroelastic crystal KSCN very close to $T_c$. One can see regions, where the needles have already formed (right side) as well as junctions of opposite sign attracting each other to form the needle tips when the crystal will be cooled to lower temperatures.

In a beautiful work Torrés, et al.[35] have investigated the interactions between ferroelastic domain walls using dislocation theory. They have shown, that needle shaped domains can be described by an effective elastic dipole, which produces long range stresses, that can *stabilize* an array of ferroelastic domains.

In the following we will use the corresponding free energy to calculate the elastic susceptibility of a multidomain crystal. In response to an appropriate applied external static and dynamic stress σ=σ$_{stat}$ + σ$_{dyn}$ the width x$_+$ of domains with ε$_s$(+) enlarges, whereas the width x$_-$ with ε$_s$(-) shrinks (Fig. 5). This leads to a new period under applied stress, i.e. $x_0(\sigma)=\frac{1}{2}[x_+(\sigma)+x_-(\sigma)]$. The macroscopic strain ε$^{DW}$ due to the domain wall motion under applied stress is then

$$\varepsilon^{DW} = \varepsilon_s \left(1 - \frac{x_-}{x_0}\right) \tag{4}$$

and the domain wall contribution to the elastic compliance can be written as

$$\Delta S^{DW} = \frac{\partial \varepsilon^{DW}}{\partial \sigma} = -\frac{\varepsilon_s}{x_0} \frac{\partial x_-}{\partial \sigma} \tag{5}$$

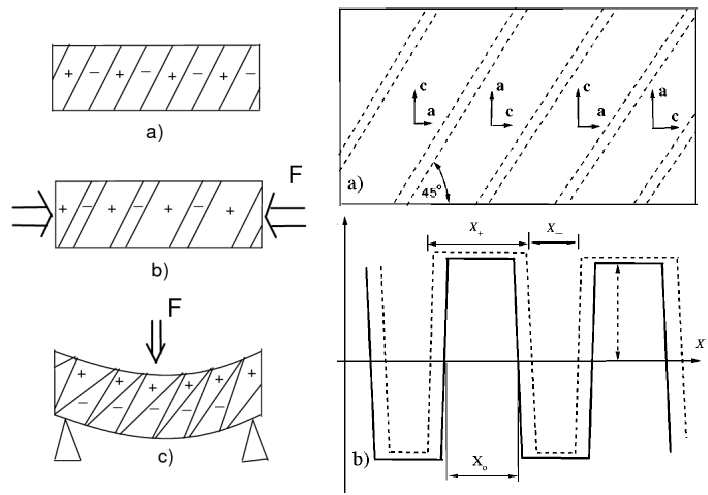

**Figure 5**. Schematic representation of ferroelastic domains and their distortion under an applied external force F.

Eq.(5) shows the main ingredients needed for the description of domain wall induced elastic effects. These are the spontaneous strain ε$_s$, the average distance of domain walls x$_0$ (or equivalently the number of domain walls N~1/x$_0$) and the change of the displacement x$_-$ due to the applied stress, i.e. the inverse effective spring constant k. In the simplest model the relation $\frac{\partial x_-}{\partial \sigma} \propto \frac{\varepsilon_s}{k}$ holds, which leads to

$$\Delta S^{DW} \propto \frac{\varepsilon_s^2}{x_0 k} \tag{6}$$

This relation is very similar to the one obtained for the contribution of ferroelectric domains to the dielectric permittivity[12] where ε$_s$ is replaced by the spontaneous polarization P$_s$. However, since the presently considered perovskites are improper ferroelastic where $\varepsilon_s \propto \eta^2$, Eq.(6) leads to $\Delta S^{DW} \propto \frac{\eta^4}{x_0 k}$ in sharp contrast with the empirical relation of Eq.(3), i.e. $\Delta S^{DW} \propto \eta^2$ (note, that the average domain width $x_0$ was experimentally found to be temperature independent, and also k is temperature independent in this simple model).

To overcome this problem we recall, that this simple model – like most of the theories of domain wall motion – assumes zero thickness of domain walls. However, it has been shown[26] that ferroelastic domain walls are rather thick, i.e. increasing from few lattice constants far below T$_c$ up to tens of lattice constants just below T$_c$. Taking into account the repulsion between ferroelastic domain walls of finite thickness Eq.(6) changes to[18]

$$\Delta S^{DW} \propto \frac{\varepsilon_s^2 w^2}{x_0 e^{-x_0/w}} \quad (7)$$

Since for a 2-4 Landau potential[26] $w^2 \propto \frac{1}{\eta^2}$ – and $x_0 e^{-x_0/w}$ was shown[18] to be almost temperature independent – we obtain $\Delta S^{DW} \propto \eta^2$ in perfect agreement with the experimental data on SrTiO$_3$, KMnF$_3$ (Fig.6) and LaAlO$_3$. Taking into account the stress dependence of the domain width $x_0(\sigma)$ we can also fit the stress dependence of the elastic susceptibility of multidomain perovskites[18].

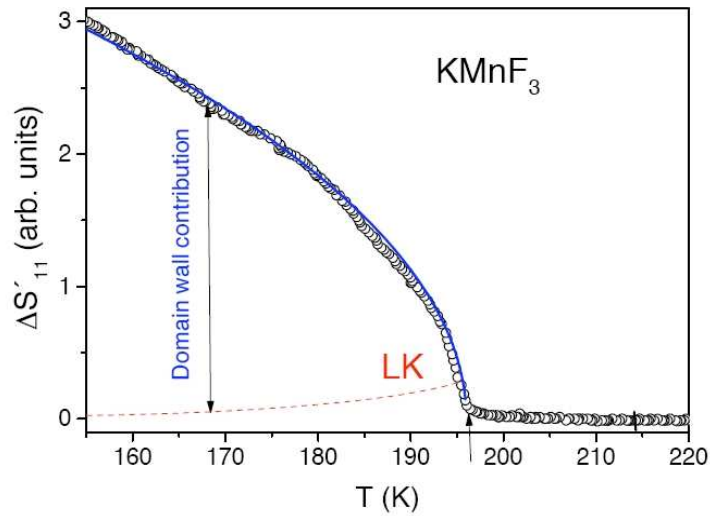

**Figure 6**. Temperature dependence of the anomalous part of the compliance $\Delta S_{11}$ of KMnF$_3$ measured by the PPS-method at f=9 Hz (points) and a fit (blue line) using Eq.(3). The dashed line marks the Landau-Khalatnikov contribution, originating from the $\eta^2\varepsilon$ coupling in the Landau free energy.

### 3.2. Proper or pseudo-proper ferroelastic phase transitions

In principle Eq.(7) should work also for proper, as well as for pseudo-proper ferroelastic phase transitions. However, in contrast to improper ferroelastic phase transitions for the proper (or pseudo-proper) case, the spontaneous strain is the primary order parameter, i.e. $\varepsilon_s \equiv \eta$ (or bilinearly coupled to the primary order parameter, i.e. $\varepsilon_s \propto \eta$). As a result the temperature dependencies of $\varepsilon_s^2$ and $w^2$ cancel each other in Eq.(7) and the temperature dependence of the domain wall motion dominated elastic compliance is mainly determined by the temperature dependence of the average domain size $x_0$ (or equivalently by the number N of domain walls), i.e.

$$\Delta S^{DW} \propto \frac{1}{x_0 e^{-x_0/w}} \propto N(T) e^{1/Nw} \tag{8}$$

Approaching a proper or pseudoproper ferroelastic phase transition from below, the number of domain walls diverges, i.e. $N \rightarrow \infty$ for $T \rightarrow T_c$. As a result the elastic compliance diverges, i.e. $\Delta S^{DW} \propto N(T) e^{1/Nw} \rightarrow \infty$ for $T \rightarrow T_c$ (note that $e^{1/Nw} \rightarrow 1$ for $T \rightarrow T_c$). This is in sharp contrast to improper ferroelastic phase transitions, where $\Delta S^{DW} \propto \eta^2 \rightarrow 0$ for $T \rightarrow T_c$.

In the following we will apply the present model to the case of lanthanum/neodymium pentaphosphate[36] $La_{1-x}Nd_xP_5O_{14}$ (LNPP). Similar as the pure compound $NdP_5O_{14}$ (NPP)[37], LNPP undergoes a proper ferroelastic second order phase transition at $T_c \approx 414$ K from monoclinic *P2₁/c* to orthorhombic *Pncm*. The primary order parameter is the shear strain $\varepsilon_5$. In the ferroelastic phase two equivalent orientational (ferroelastsic) domain states with opposite shear strain $\pm\varepsilon_5$ appear. These ferroelastic domains form a regular stripe pattern and the number of domains increases drastically when approaching $T_c$ from below[36,37]. The transverse elastic constant $C_{55}$ of $Nd_{0.9}La_{0.1}P_5O_{14}$ (LNPP) was measured by a resonator method at several tens kHz[36].

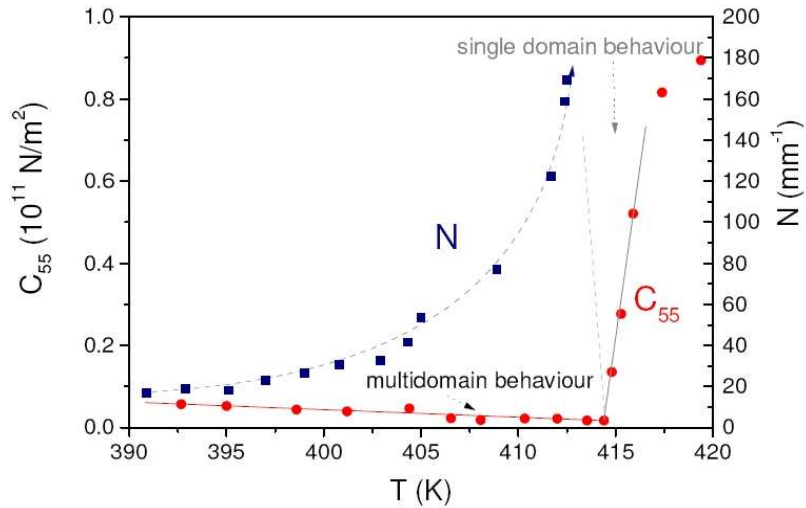

**Figure 7**. Temperature dependencies of shear elastic constant $C_{55}$ and domain wall density $N \propto 1/x_0$ of LNPP (from Ref.36). The lines are guides to the eye, except for the dashed grey ones, which show the calculated inverse Curie-Weiss behaviour

($C_{55} \propto |T - T_c|$) of a monodomain sample.

As Fig.7 shows it displays the inverse Curie-Weiss type anomaly above $T_c$=414 K as expected for a proper ferroelastic phase transition. But in the ferroelastic phase the temperature behaviour of $C_{55}$ is in contrast to the expected inverse Curie-Weiss behaviour. It stays at a very low value in a broad temperature range, i.e. increasing from ≈ 0 GPa at $T_c$ to ≈ 5 GPa at $T_c$ – 20 K. This huge softening is due to the motion of ferroelastic domain walls.

Fig.8 shows the temperature dependence of the shear compliance $S_{55} = C_{55}^{-1}$. The domain wall contribution to the shear compliance can be perfectly fitted using Eq.(8) with the measured values of N(T).

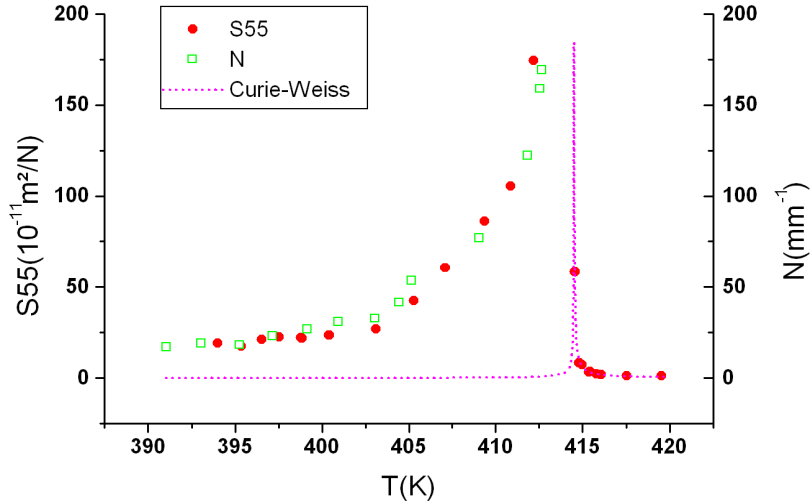

**Figure 8**. Measured temperature dependence of the shear elastic compliance $S_{55} = C_{55}^{-1}$ of LNPP (From Ref.36) fitted with Eq.(8) using the data for N(T) measured with heating.

## 4. Conclusions

We have studied the influence of domain wall motion to the elastic susceptibilities of improper and proper ferroelastic materials. A detailed theoretical analysis shows that the domain wall contribution to the elastic compliance $\Delta S^{DW}$ is very different for both classes of materials. For improper ferroelastics it is proportional to the square of the order parameter which vanishes at $T_c$, i.e. $\Delta S^{DW} \propto \eta^2(T) \to 0$ for T→$T_c$, while for proper ferroelastics it scales with the number of domain walls N which diverges at $T_c$, i.e. $\Delta S^{DW} \propto N(T) \to \infty$ for T→$T_c$. Comparison with numerous experimental data on perovskites (improper ferroelastic) and La$_{1-x}$Nd$_x$P$_5$O$_{14}$ (proper ferroelastic) yields perfect agreement.

**Acknowledgements**

W.S. would like to use this opportunity to thank Wolfgang Kleemann for his friendship and innumerable stimulating discussions which started almost 20 years ago and that have been continuing till now.